\documentclass[iop, jcp, twocolumn, floatfix, superscriptaddress, reprint, 10pt]{revtex4-1}
\usepackage{extsizes}
\usepackage[paperwidth=210mm,paperheight=297mm,centering,hmargin=2cm,vmargin=2.5cm]{geometry}
\usepackage[utf8]{inputenc}

\usepackage{verbatim}
\usepackage{bm}
\usepackage{amsmath}
\usepackage{amsfonts}
\usepackage{natbib}
\usepackage{physics}

\usepackage{pdfpages}
\usepackage{csvsimple}
\usepackage{graphics}
\usepackage{float}
\usepackage[capitalize]{cleveref}
\usepackage{multirow}
\usepackage{booktabs}
\usepackage{makecell}

\newcommand{\xj}{\mathcal{X}_{j}}
\newcommand{\xk}{\mathcal{X}_{k}}
\newcommand{\br}{\mathbf{r}}
\newcommand{\bu}{\mathbf{u}}
\newcommand{\Rhat}{{\hat{R}}}
\newcommand{\bk}{\mathbf{k}}

\newcommand{\bt}{\mathbf{t}}

\newcommand{\dbr}{\textrm{d}\br}
\newcommand{\dint}{\textrm{d}}
\newcommand{\drhat}{\textrm{d}\hat{R}}

\newcommand{\CX}{\mathcal{X}}

\newcommand{\CA}{\mathcal{A}}
\newcommand{\CB}{\mathcal{B}}

\newcommand{\iket}[2]{\ket{#1}_{#2}}
\newcommand{\ibraket}[3]{\braket{#1}{#2}_{#3}}

\makeatletter
\AtBeginDocument{\let\LS@rot\@undefined}
\makeatother

\begin{document}

\setcitestyle{super}

\title{%
Atom-Density Representations for Machine Learning}
\author{Michael J. Willatt}
\affiliation{Laboratory of Computational Science and Modeling, Institute of Materials, \'Ecole Polytechnique F\'ed\'erale de Lausanne, 1015 Lausanne, Switzerland}
\author{F\'elix Musil}
\affiliation{Laboratory of Computational Science and Modeling, Institute of Materials, \'Ecole Polytechnique F\'ed\'erale de Lausanne, 1015 Lausanne, Switzerland}
\affiliation{National Center
for Computational Design and Discovery of Novel Materials (MARVEL)}
\author{Michele Ceriotti}
\email{michele.ceriotti@epfl.ch}
\affiliation{Laboratory of Computational Science and Modeling, Institute of Materials, \'Ecole Polytechnique F\'ed\'erale de Lausanne, 1015 Lausanne, Switzerland}

\onecolumngrid
\begin{abstract}
The applications of machine learning techniques to chemistry and materials science become more numerous by the day. The main challenge is to devise representations of atomic systems that are at the same time complete and concise, so as to reduce the number of reference calculations that are needed to predict the properties of different types of materials reliably. This has led to a proliferation of alternative ways to convert an atomic structure into an input for a machine-learning model. We introduce an abstract definition of chemical environments that is based on a smoothed atomic density, using a bra-ket notation to emphasize basis set independence and to highlight the connections with some popular choices of representations for describing atomic systems. 
The correlations between the spatial distribution of atoms and their chemical identities are computed as inner products  between these feature kets, which can be given an explicit representation in  terms of the expansion of the atom density on orthogonal basis functions, that is equivalent to the smooth overlap of atomic positions (SOAP) power spectrum, but also in real space, corresponding to $n$-body correlations of the atom density.   
This formalism lays the foundations for a more systematic tuning of the behavior of the representations, by introducing operators that represent the correlations between structure, composition, and the target properties. It provides a unifying picture of recent developments in the field and indicates a way forward towards more effective and computationally affordable machine-learning schemes for molecules and materials.
\end{abstract}
\twocolumngrid

\maketitle

\section{Introduction}

Machine learning (ML) is used routinely nowadays as an expedient to circumvent costly electronic structure calculations. Given a set of atomic structures and a means to represent them concisely as a  vector of numbers (often referred to as features, descriptors or fingerprints), machine learning models can make property predictions by interpolating over the known properties of a subset used for training.
Provided the machine learning model yields an acceptable level of accuracy, this allows for exploratory investigations of vast numbers of molecules in chemical-compound space, which is essential for e.g. high-throughput screening, drug discovery and molecular classification.\cite{Ramakrishnan2015,Faber2016,Bartok2017,Montavon2013,Mitchell2014,Carrete2014,Ward2016} 
It also enables the development of atomic potentials with the accuracy of \textit{ab initio} alternatives with a smaller computational expense.\cite{behl-parr07prl, Szlachta2014, Bartok2010,Morawietz2012,Kobayashi2017,Smith2017,Khorshidi2016,Gastegger2017,Glielmo2018,Chmiela2017,Yao2017}

For a ground-state electronic structure calculation, an atomic structure is completely characterized by the number of electrons, the positions of the nuclei and their identities (nuclear charges), and in principle a machine learning model could use the same information as a representation of its input. However, this information is not consistent with obvious physical symmetries. Any translation or rotation of a structure, or permutation of identical atoms within, will not affect scalar properties like energies. A representation that reflects this invariance is therefore desirable for an efficient comparison to be made between structures.\cite{Bartok2013}

Different strategies have been proposed to incorporate these symmetries. Approaches based on internal coordinates (e.g. Coulomb matrices~\cite{Rupp2012,Montavon2013,Barker2017}, eigenvalues of overlap matrices~\cite{Sadeghi2013} or deep potential molecular dynamics~\cite{Zhang2018}) are automatically invariant to rotations and translations but require an additional symmetrization over the permutation group. For low-dimensional problems this symmetrization can be performed exactly~\cite{braa-bowm09irpc,Xie2010,Jiang2013}. For larger systems, one can proceed by sorting the vector of interatomic distances or eigenvalues of a matrix that depends on interatomic distances.~\cite{Sadeghi2013} However, both procedures introduce derivative discontinuities. 
Instead, many approaches to represent atomistic configurations, e.g radial distribution functions~\cite{Schutt2014}, wavelets~\cite{Hirn2017,Eickenberg2017}, invariant polynomials~\cite{Shapeev2015},  symmetry functions following Behler and Parrinello~\cite{behl-parr07prl,Smith2017}, or a many-body tensor representation~\cite{Zhang2018}, rely more or less explicitly on atomic distributions. 

In this article, we start from an abstract representation of structures and atomic environments to discuss atom-density-based approaches to chemical machine learning. We emphasize the basis-set independence of this representation by using the Dirac bra-ket notation. 
This framework provides a unifying picture of the field, in that several  popular techniques can be seen as alternative representations of the same abstract feature vectors. 
In particular, we show that by representing these kets based on an expansion of atom-centered Gaussians in radial basis functions and spherical harmonics one recovers the SOAP power spectrum, which has been used very successfully to compare structures and predict scalar properties\cite{Bartok2013a, Szlachta2014, Deringer2017,Jinnouchi2017,PhysRevB.90.104108,Deringer2017,Kajita2017}, has been generalized to model tensorial  properties \cite{Grisafi2018} and sparsified to reduce the computational cost \cite{imba+18jcp}.
This formalism also provides a way to fine-tune or reduce the dimensionality of the feature vector by introducing a general coupling between different components in the definition of the kernel, recovering the flexibility of using different kinds of density-based representations within an elegant, unified framework.

\section{A Dirac notation for atomic  configurations and environments}

The Dirac (bra-ket) notation is often used to streamline the formulation of quantum-mechanical expressions, since it stresses basis-independence of quantum states. 
In a kernel ridge (Gaussian process) regression framework, kernel functions between atomic configurations $K(\CA,\CB)$ are used as the basis to build machine-learning models of atomic-scale properties
\begin{equation}
y(\CA) = \sum_M x_M K(\CA,\CX_M).
\end{equation}
Here $y$ is a property one wants to predict, $\CX_M$'s are a set of reference atomic structures, and $x_M$ are weights that can be determined by minimization of a loss function that measures the discrepancy between the predictions of the model and a set of reference calculations.

Well-behaved (positive-semidefinite) kernels correspond to inner products between (possibly infinite-dimensional) feature vectors, and are also independent of the choice of basis that is used to represent them.
This suggests that it might be beneficial to adopt the bra-ket notation to formulate input representations $\ket{\CA}$ and the associated kernels $K(\CA,\CB)=\bra{\CA}\ket{\CB}$. When using feature vectors explicitly in a computer simulation, it is necessary to settle on some (finite) basis. In combination with the advantage that some bases provide over others for proving certain results, this is perhaps why the Dirac notation has not been used in this context up to now.

\subsection{Density-based structural representations}

As a starting point to represent atomic structures, every atomic configuration $A$ could be associated with a ket $\ket{A}$ that describes the elemental composition and geometric arrangement of atoms. To make this idea concrete, we might center a smooth, real, positive function $g(\br)$ (e.g. a normalized Gaussian) on each atom and decorate them with orthonormal kets $\ket{\alpha}$ to represent their elemental identities. Smoothness in the representation is beneficial as it leads to smooth kernels and better-behaved regression~\cite{Rasmussen2006}. 
Such an atomic configuration is written in position space as
\begin{equation}
    \bra{\br}\ket{A} = \sum_i g(\br -\br_i) \ket{\alpha_i},
    \label{eq:dirac-a}
\end{equation}
where the sum is taken over all atoms in the configuration. This expansion could be generalized by using e.g. element-dependent widths in $g(\mathbf{r})$, i.e. $g(\mathbf{r}) \to g(s(\alpha)\mathbf{r})$. 
Regardless of the particular form of $g(\mathbf{r})$ (provided that it is sufficiently localized),
$\ket{A}$ provides a unique representation of the structure, but is variant with respect to fundamental physical symmetries, such as rigid translations $\hat{t}$ and rotations $\hat{R}$ of the constituent atoms $\{\mathbf{r}_{i}\} \to \{\hat{R}\hat{t}  \mathbf{r}_{i} \}$. As has been stressed many times previously, to achieve efficient learning one should use representations that possess the same symmetries as the property one wants to learn.\cite{behl-parr07prl,Bartok2013,glie+17prb,Grisafi2018}

\subsection{Symmetry-invariant representations}

To address the variance of \cref{eq:dirac-a} with respect to a symmetry operation $\hat{S}$, one can proceed by formally averaging the ket over the corresponding symmetry group (a procedure often referred to as Haar integration\cite{nachbin1976haar}):
\begin{equation}
\iket{\CA^{(1)}}{\hat{S}} = \int \dint\hat{S}\, \hat{S}\ket{\CA}.
\end{equation}
To see how this translates into symmetry invariant representations, let us start by considering the relatively simple case of the integration over the translation group which simply corresponds to the integration over $\mathbb{R}^3$.
Averaging directly over the position representation of $\ket{A}$ leads to a rather uninformative representation, which eliminates all structural information and only counts the number $N_\alpha$ of atoms belonging to each species,
\begin{equation}
\begin{split}
\ibraket{\br}{\CA^{(1)}}{\hat{t}}=&\int \dint \hat{t} \bra{\br}\hat{t}\ket{\CA} \\
=&\sum_i \ket{\alpha_i} \int\dint {\bt}\, g(\bt + \br -\br_i)= \sum_\alpha N_\alpha \ket{\alpha},
\end{split}
\label{eq:a-haar-1}
\end{equation}
where we have used the position representation of the translation operator.
To avoid this information loss, one can perform the Haar integration over tensor products of $\ket{A}$, and define
\begin{equation}
\iket{\CA^{(\nu)}}{\hat{t}} =
\int \dint \hat{t}
\underbrace{\hat{t}\ket{\CA}\otimes\hat{t}\ket{\CA} \ldots \hat{t}\ket{\CA}}_\nu.
\label{eq:haar-t-nu}
\end{equation}
For $\nu=2$, and assuming for simplicity that the same smooth density function is used for each atom, one gets
\begin{equation}
\begin{split}
\ibraket{\br \br'}{\CA^{(2)}}{\hat{t}} = &
\int \dint \hat{t} \sum_{ij} g(\hat{t} \br - \br_i) g(\hat{t} \br' - \br_j)
\ket{\alpha_i \alpha_j} \\
=& \sum_{ij} \ket{\alpha_i \alpha_j} \int \dint \bt \,g(\bt + \br - \br_i) g(\bt + \br' - \br_j) \\
=& \sum_{ij} \ket{\alpha_i\alpha_j}(g * g)(\br'-\br - \br_{ij}),
\end{split}
\label{eq:conv}
\end{equation}
where $\br_{ij}=\br_i-\br_j$, and $*$ is the standard convolution operator. 
We can simplify the notation for $\iket{\CA^{(2)}}{\hat{t}}$ in the position representation by (1) noting that the convolution in \cref{eq:conv} only depends on $\br-\br'$, so we can write the ket as a function of $\Delta\br=\br-\br'$ alone;
(2) redefining the convolution of two atom-density functions as $h=g * g$:\footnote{For a generic basis function $g*g$ might be a complicated function, but when $g$ is a Gaussian, $g*g$ is simply a Gaussian with double the variance}
\begin{equation}
\ibraket{\Delta\br}{\CA^{(2)}}{\hat{t}} =
\sum_j \ket{\alpha_j} \sum_i h(\Delta\br-\br_{ij}) \ket{\alpha_i}.
\label{eq:r-at2}
\end{equation}

\subsection{Tensor-product representations}

Before proceeding further, let us comment briefly on the implications of the form of this ket for machine-learning of the properties associated with the structure $\ket{A}$ using kernel ridge regression, taking for simplicity a single-species compound so we can ignore the elemental kets.
Learning from a linear kernel $K(\CA,\CB)=\ibraket{\CA^{(2)}}{\CB^{(2)}}{\hat{t}}$ is equivalent to the optimization of a linear mapping between the ket and the property, i.e.
\begin{equation}
y(\mathcal{A}) = %
\int \dint \br \,
y(\br) \ibraket{\br}{\CA^{(2)}}{\hat{t}}.
\end{equation}
Taking the Dirac $\delta$ distribution limit of $g(\mathbf{r})$, one sees this is a (orientation-dependent) pair potential,
\begin{equation}
y(\mathcal{A}) = \sum_{ij} y(\br_{ij}),
\end{equation}
and it is therefore easy to conceive of properties that cannot be represented in this form.
The feature vector itself, however, contains complete information about the structure, which can be recovered by taking tensor products of $\ket{\CA}$, or (equivalently) raising the associated kernel to an integer power. For instance, if one takes the outer product of the translationally-symmetrized ket (whose corresponding kernel is just the elementwise square of the linear kernel), learning amounts to the optimization of a function that depends on two displacement vectors simultaneously,
\begin{equation}
\begin{split}
y(\mathcal{A})=&
 \int \dint \br \dint \br' 
y(\br,\br')
\ibraket{ \br'}{\CA^{(2)}}{\hat{t}} \ibraket{ \br}{\CA^{(2)}}{\hat{t}}  \\
=&  \sum_{iji'j'} 
\int \dint \br \dint\br' y(\br,\br') h(\br -\br_{ij}) h(\br' - \br_{i'j'})
 \\
 \underset{h\rightarrow\delta}{=}  & \sum_{iji'j'} y(\br_{ij},\br_{i'j'}),
\end{split}
\end{equation}
and so on. 
This simple example highlights how high-order correlations between atomic positions can be incorporated in the model by taking the tensor product of the structural ket before taking the Haar integral~\cite{Bartok2013,Glielmo2018} (that is, choosing a high value of $\nu$ in Eq.~\eqref{eq:haar-t-nu}) or by taking a tensor product of the invariant ket, 
\begin{equation}
    \underbrace{\ket{\CA^{(\nu)}}_{\hat{t}}\otimes \ket{\CA^{(\nu)}}_{\hat{t}}\otimes \dots\otimes \ket{\CA^{(\nu)}}_{\hat{t}}}_{{\zeta}} \to \ket{\CA^{(\nu)}}_{\hat{t}}^{(\zeta)}.
    \label{eq:general-zeta}
\end{equation}
The latter choice corresponds to taking elementwise powers of the linear invariant kernel. In other terms, using a \emph{unique} representation of a structure in a non-linear ML model can introduce higher body order correlations than those explicitly afforded by the feature vector itself. 

\subsection{Atom-centered representations} 

Having clarified how tensor-product kets can be used to incorporate higher-order correlations between the atoms, let us move on to discuss how an atom-centered description arises naturally as a by-product of symmetrization over the translation group.  By grouping together the terms in the sum corresponding to displacement vectors involving atom $j$,  the translationally-invariant second-order ket Eq.~\eqref{eq:r-at2}  decomposes into atom-centered contributions, 
\begin{equation}
\iket{\CA^{(2)}}{\hat{t}} = \sum_j \ket{\alpha_j}\ket{\xj},
\label{eq:environs}
\end{equation}
where we have dropped the indication of the translational averaging from $\ket{\xj}$ to keep at bay the complexity of the notation.
Note that Eq.~\eqref{eq:environs} implies an additive definition of the relation between the representations of the entire structures, and those associated with atom-centered environments. This definition translates into a ``global'' kernel between structures that is a sum (or average) of kernels between environments. While other choices are possible~\cite{De2016}, this formal derivation shows how naturally an additive kernel arises from a symmetrization of a density-based ``global'' feature vector.

The position representation of the environmental atom-centered ket $\ket{\xj}$ is 
\begin{equation}
    \bra{\br}\ket{\xj} = \sum_i f_c(r_{ij})
    h(\br-\br_{ij})\ket{\alpha_i}.
    \label{eq:r-xj}
\end{equation}
In this definition we have introduced a smooth cutoff function $f_c(r_{ij})$ so that each environment only depends on the position of the atoms in a spherical neighborhood centered on atom $j$. 
While one could in principle proceed with an atom-centered description that incorporates information from the entire structure, by making $f_c(r)=1$, localisation is useful for computational reasons and is justified when studying atomic problems in light of the nearsightedness principle of electronic matter\cite{prod-kohn05pnas}, which underlies most linear-scaling electronic structure methods~\cite{yang91prl,gall-parr92prl,goed99rmp,ceri+08jcp}.
Note that when the ket is written in this form it might make sense to further generalize the definition of $h$, e.g. by making its width dependent on $r_{ij}=|\br_{ij}|$, $h(\br) \to h(s(r_{ij})\br)$, or by choosing a form other than a Gaussian that is more flexible or computationally efficient. 
The notation can be further simplified by emphasizing the representations of structure and composition,
\begin{equation}
 \psi^\alpha_{\xj}(\br) \equiv  \bra{\alpha \br}\ket{\xj} = \sum_{i \in \alpha} f_c(r_{ij})
    h(\br-\br_{ij}).
    \label{eq:ar-xj-psi}
\end{equation}
Writing the ket as a sum over all elements $\alpha = \textrm{H}, \textrm{He}, \dots$
\begin{equation}
    \bra{\br}\ket{\xj} = \sum_\alpha \psi^\alpha_{\xj}(\br)\ket{\alpha}.
    \label{eq:r-xj-psi}
\end{equation}

This translationally-invariant atom-centered environment representation can also be adapted by taking a linear transformation $\hat{U}\ket{\xj} \to \ket{\xj}$, where the linear operator $\hat{U}$ might act in the position space, the element space or both. 
As we will see, the freedom in choosing the form of $\hat{U}$ can be used to tune the behavior of the representation to describe in a more efficient way the relation between structure and properties. 

\subsection{Rotationally-invariant representations}

In order to obtain a rotationally-invariant representation, one can formally average the ket $\ket{\CX}$ over the $SO(3)$ rotation group,
\begin{equation}
     \iket{\CX^{(1)}}{\hat{R}} =
     \int d\hat{R}\, \hat{R} \ket{\CX}.
    \label{eq:raw-1}
\end{equation}
This ket can be readily computed in the position representation. Taking for simplicity the case where only one element is present, one gets
\begin{equation}
\ibraket{\br}{\CX^{(1)}}{\hat{R}} =\int \dint \hat{R}\, \psi(\hat{R} \br) =  \int \dint \hat{R}\, \psi(r \hat{R} \hat{e}_z),
\label{eq:rot-nu1}
\end{equation}
where we have used the fact that the integral is over all the rotation matrices, and so we can always rotate $\br$ to be aligned with the Cartesian $z$ axis $\hat{e}_z$ before taking the integral. 
The average can be written explicitly in terms of a suitable parameterization of the rotations, e.g. using Euler angles,
\begin{equation}
\frac{1}{8\pi^2} \int_0^{2\pi} \dint \alpha \int_0^\pi \sin\beta \dint\beta \int_0^{2\pi} \dint \gamma \,\psi(r\hat{R}(\alpha,\beta,\gamma) \hat{e}_z) .
\end{equation}
One can then recognize that the $\gamma$ angle does not affect $\hat{e}_z$, so the integral can be written equivalently as an average over the unit sphere\footnote{This is a consequence of the fact that $SO(3)$ is the product of $SO(2)$ and $S^2$}. 
We can then define
\begin{equation}
\ibraket{r}{\CX^{(1)}}{\hat{R}} \propto r \int \dint \hat{R}\, \psi(r \hat{R} \hat{e}_z) = \frac{1}{4\pi} r \int \dint \hat{\br}\, \psi(r\hat{\br}),
\label{eq:so3-param}
\end{equation}
where we have highlighted the fact that the position representation only depends on $r$, and we have explicitly included a factor of $r$ so that 
\begin{equation}
\int \dint\br 
\ibraket{\xk^{(1)}}{\br}{\hat{R}}
\ibraket{\br}{\xj^{(1)}}{\hat{R}}=\int \dint r 
\ibraket{\xk^{(1)}}{r}{\hat{R}}
\ibraket{r}{\xj^{(1)}}{\hat{R}}.
\end{equation}

Much like in the case of translations, the average over rotations eliminates too much information, and $\iket{\CX^{(1)}}{\hat{R}}$ does not retain knowledge of the angular correlations of atoms around the center of the environment. 
A more general family of invariant kets can be obtained by starting from the tensor products of (possibly different) environmental kets, $\hat{U}_1\ket{\CX^1}\otimes \hat{U}_2\ket{\CX^2}\otimes\ldots$, and then symmetrizing over the rotation group,
\begin{equation}
\iket{\CX^{(\nu)}}{\hat{R}} =
\int \dint\hat{R} \, \prod_{\aleph}^{\nu} \, \otimes \, \hat{R}\hat{U}_{\aleph} \ket{\xj^{\aleph}}.
    \label{eq:raw-nu}
\end{equation}  

As for the case of translational averages, one can use a linear map (or equivalently a linear kernel) to build a machine-learning model of a property based on these symmetrized kets. 
Non-linear kernels correspond to tensor products of symmetrized kets such as
\begin{equation}
\underbrace{\iket{\CX^{(\nu)}}{\hat{R}}\otimes \iket{\CX^{(\nu)}}{\hat{R}}\otimes \ldots \otimes \iket{\CX^{(\nu)}}{\hat{R}}}_{{\zeta}} \to \iket{\CX^{(\nu)}}{\hat{R}}^{(\zeta)},
    \label{eq:soap-zeta}
\end{equation}
and one could further generalize the construction by taking products of kets built from different $\hat{U}$ operators.

\begin{figure*}[bhtp]
\centering
\includegraphics[width=1.0\linewidth]{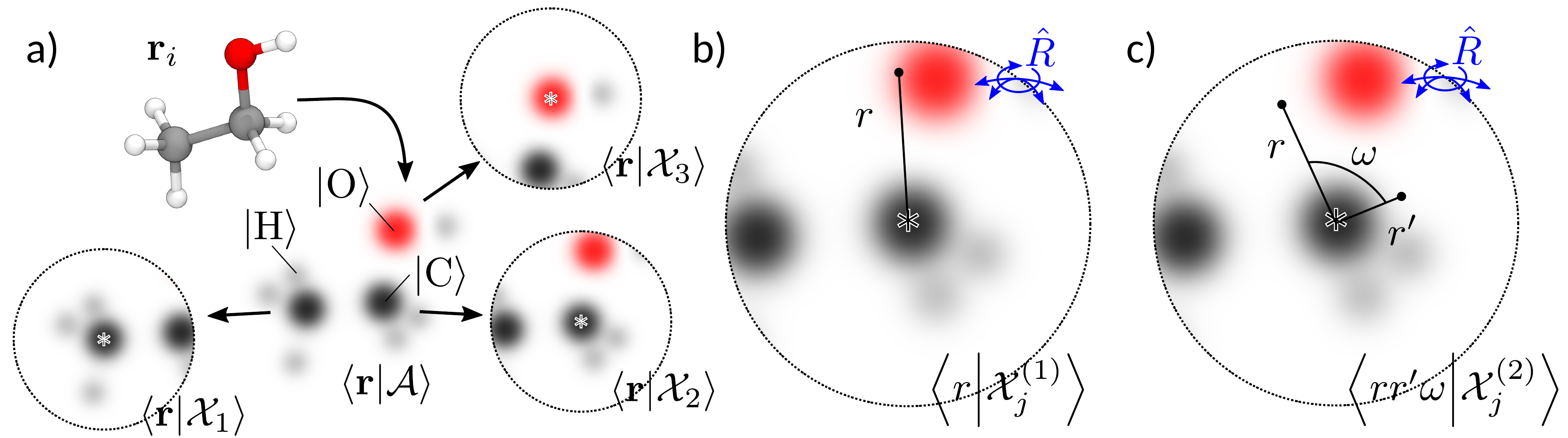}
\caption{Atom-density-based structural representations, expressed in the real-space $\bra{\br}$ basis. 
(a) A structure can be mapped onto a smooth atom density built as a superposition of smooth atom-centered functions. The overall density can be decomposed in atom-centered environments, and information on chemical compositions can be stored by decorating the functions with elemental kets.
(b) The $\nu=1$ invariant ket corresponds to spherical averaging of the environmental atom density. (c) The $\nu=2$ invariant ket corresponds to three-body correlations, which are obtained by integrating over all rotations a stencil corresponding to two distances along two directions with a fixed angle $\arccos \omega$ between them.}
\label{fig:soap-realspace}
\end{figure*}

\section{A unified picture of density-based representations}

\cref{eq:raw-nu} provides an very general -- and abstract -- definition of a density-based representation of an atomic structure that encodes translational, rotational and permutation symmetries.
This level of abstraction provides a unifying picture of the field, in that many of the representations that have been used for machine-learning of atomic-scale properties can be seen as special cases of this form, or as the result of projection onto a particular choice of basis.

Let us start by considering the translationally-invariant ket $\iket{\CA^{(2)}}{\hat{t}}$, writing it in a plane-waves basis $\{\ket{\mathbf{k}}\}$, and taking for simplicity the $h\rightarrow\delta$ limit. One obtains a representation that is equivalent to the diffraction pattern generated by the structure, decomposed in multiple channels that correspond to the reciprocal-space correlations between different atomic species,
\begin{equation}
\ibraket{\bk}{\CA^{(2)}}{\hat{t}} = \sum_{ij} \ket{\alpha_i\alpha_j} e^{\mathrm{i} \bk \cdot \br_{ij}}.
\end{equation}
When considering a periodic structure, and with an appropriate normalization, this representation is directly connected with the fingerprints that have been recently used to identify crystalline structures~\cite{zile+18nc}, highlighting how different choices of basis may be best suited to different applications. 

\subsection{Many-body kernels and representations}
Moving on to the case of rotationally-invariant kets, let us take for simplicity $\hat{U}_{\aleph}=1$, and assume that all the environmental kets that are multiplied in \cref{eq:raw-nu} are the same. We will revisit later the possibility of introducing a linear operator to fine-tune the properties of the representation. Since we have started from a position representation for the environmental kets, it is natural to write \cref{eq:raw-nu} explicitly in a complete basis of position and element states, $\ket{\prod_{\aleph}^{\nu}\alpha_{\aleph}\mathbf{r}_{\aleph}} \equiv \prod_{\aleph}^{\nu}\otimes \ket{\alpha_{\aleph}\mathbf{r}_{\aleph}}$,
\begin{equation}
    \bra{\prod_{\aleph}^{\nu}\alpha_{\aleph}\mathbf{r}_{\aleph}}\ket{\mathcal{X}_{j}^{(\nu)}}_\Rhat = \int d\hat{R} \prod_{\aleph}^{\nu} \bra{\alpha_{\aleph}\hat{R} \mathbf{r}_{\aleph}}\ket{\mathcal{X}_{j}}.
    \label{eq:desc_in_pos}
\end{equation}

One can see clearly that the kernels associated with Eq.~\eqref{eq:desc_in_pos} are in the form of the invariant $n$-body kernels discussed in Ref.~\citenum{Glielmo2018} (more specifically, as we will see below, they correspond precisely to the SOAP kernels if $g$ is a Gaussian). Considering the case with a single element,
\begin{equation}
\ibraket{\xk^{(\nu)}}{\xj^{(\nu)}}{\hat{R}}=
\int\dint \hat{R}\dint\hat{R}'\left[
\int\dint \br\, \psi_{\xk}(\hat{R}'\br)\psi_{\xj}(\hat{R}\br)
\right]^\nu,
\label{eq:haar-rotation-kernel}
\end{equation}
it is clear that one of the two Haar integrals is redundant and can be eliminated.

Let us consider the effect of $\nu$ on the representation and the information that it captures.
As discussed in the case of $\nu=1$ following Eq.~\eqref{eq:rot-nu1}, one of the input vectors $\br$ can be aligned with a fixed reference axis, e.g. $\hat{e}_z$. The fact that this axis is invariant under one of the Euler rotations makes it possible to align a second vector so that it lies in the $xz$ plane. 
For $\nu = 1$ and $\nu=2$ this analysis leads to
\begin{equation}
\begin{split}
\ibraket{\alpha r}{\CX^{(1)}}{\hat{R}} \propto &\, r \int \drhat\, \psi^\alpha(r\hat{R}\hat{e}_z) \\
\ibraket{\alpha r \alpha'r'\omega}{\CX^{(2)}}{\hat{R}} \propto &\, r r' \int \drhat\, \psi^{\alpha'}(r'\hat{R}\hat{e}_z) \\ \times &\,\psi^\alpha\left(r\hat{R}(\omega\hat{e}_z + \sqrt{1-\omega^2}\hat{e}_x)\right),
\end{split}
\label{eq:so3-realspace}
\end{equation}
where $\omega = \hat{\mathbf{r}}\cdot \hat{\mathbf{r}}'$ (see \cref{fig:soap-realspace}).
After one has aligned the first two $\mathbf{r}_\aleph$'s, the position of all the other $\mathbf{r}_\aleph$'s cannot be manipulated, so in practice for $\nu>2$ each further order brings in three degrees of freedom, that are expressed in the reference system in which the first two vectors are aligned along the $z$ axis and lie in the $xz$ plane.
For $\nu=3$,
\begin{equation}
\begin{split}
& \ibraket{\alpha r \alpha'r' \omega \alpha'' r'' \hat{\mathbf{r}}''}{\CX^{(2)}}{\hat{R}} \\ 
\propto & \, r r' r'' \int \drhat\, \psi^{\alpha}(r\hat{R}\hat{e}_z)\\ 
& \times \,\psi^{\alpha'}\left(r'\hat{R}(\omega\hat{e}_z + \sqrt{1-\omega^2}\hat{e}_x)\right) \\
& \times \,\psi^{\alpha''}\left(r''\hat{R}\hat{\mathbf{r}}''\right).
\end{split}
\label{eq:bispectrum-realspace}
\end{equation}
Also note that we have incorporated the square root of the Jacobian in the definition of the representations so that the corresponding kernels can be computed straightforwardly as the inner product between two vectors without scaling.

\begin{figure*}[bhtp]
\centering
\includegraphics[width=1.0\linewidth]{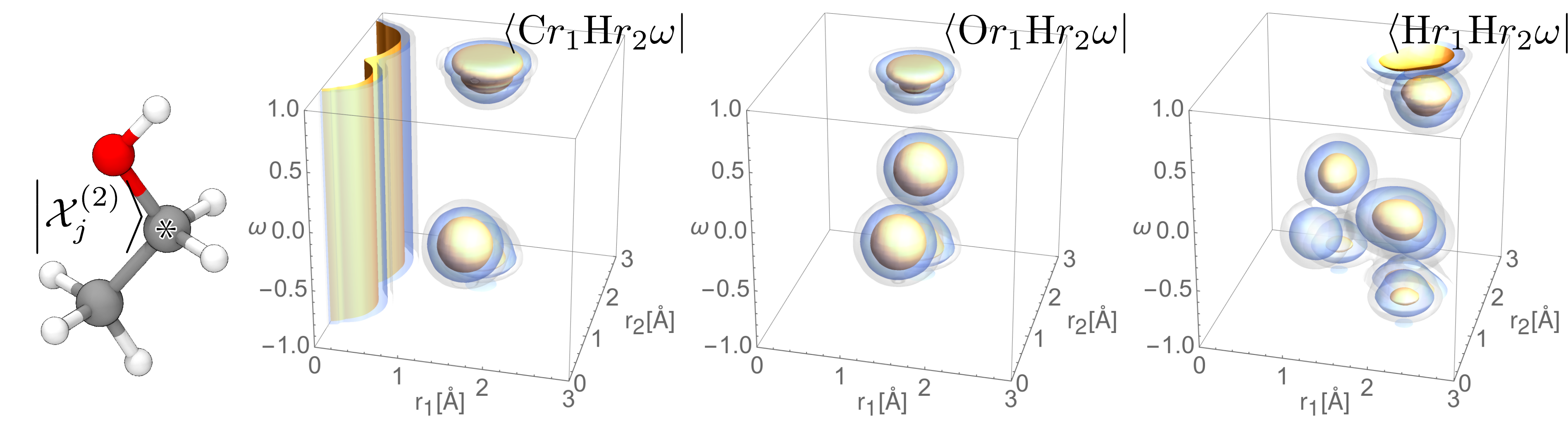}
\caption{Isocontours of the 3-body correlation functions associated with the environment centered on the tagged carbon atom of an ethanol molecule. From left to right, the figures correspond to $\bra{\text{C}r\text{H}r'\omega}\ket{\CX^{(2)}_j}_\Rhat/rr'$, $\bra{\text{O}r\text{H}r'\omega}\ket{\CX^{(2)}_j}_\Rhat/rr'$, $\bra{\text{O}r\text{H}r'\omega}\ket{\CX^{(2)}_j}_\Rhat/rr'$.
}\label{fig:3body}
\end{figure*}

By expanding the densities as sums over atoms, it becomes clear that these kets are representations of the $(\nu+1)$-body order correlations between atoms within an environment\cite{Bartok2013,Glielmo2018} (Fig.~\ref{fig:3body}).
To start with, we return to the delta function limit of the atomic densities. In the limit in which each atomic density is represented by Dirac $\delta$ distributions, the position representations of the invariant vectors take very simple forms:
\begin{equation}
\begin{split}
\bra{\alpha r}\ket{\xj^{(1)}}_\Rhat \underset{h\rightarrow\delta}{\propto}  & r \sum_{i\in \alpha} f_c(r_{ij}) \delta(r-r_{ij})\\
\bra{\alpha r \alpha' r' \omega}\ket{\xj^{(2)}}_\Rhat \underset{h\rightarrow\delta}{\propto} & r r' \sum_{i\in \alpha} \sum_{i'\in \alpha'}
 \delta(r-r_{ij})\delta(r'-r_{i'j}) \\
& \times \delta(\omega-\hat{\br}_{ij}\cdot\hat{\br}_{i'j} ) f_c(r_{ij}) f_c(r_{i'j}).\\
\end{split}  
\label{eq:delta-12}
\end{equation}
One then sees how linear regression based on $\iket{\CX^{(\nu)}}{\hat{R}}$ corresponds to $(\nu+1)$-body potentials e.g. for the 3-body term,
\begin{equation}
\begin{split}
y(\xj) =& \int \dint r \dint r' \dint\omega \,
y^{(2)}(r,r',\omega) \bra{r r' \omega}\ket{\xj^{(2)}}_\Rhat =\\
=&\sum_{ik} y^{(2)}(r_{ij},r_{jk},\omega_{ijk}).
\end{split}
\end{equation}

There are however good reasons to use non-linear functions of the feature vector in an ML model. 
In the case of sufficiently sharp atom-centered density functions, the ket with $\nu=1$ contains information on the list of all pair distances within an environment, which is not sufficient to reconstruct the structure of the environment unequivocally. The representation with $\nu=2$, on the other hand, contains information on pair distances and angles between triplets of atoms. 
To the best of our knowledge, and based on extensive numerical experiments~\cite{Bartok2013}, this information is sufficient to reconstruct a configuration modulo arbitrary rigid translations, rotations and inversion symmetry. 
Tensor products of the $\iket{\CX^{(2)}}{\hat{R}}$  ket are then sufficient as a basis to represent arbitrarily complex invariant functions of the atomic coordinates.

\subsection{Behler-Parrinello symmetry functions}
An expression of  \cref{eq:raw-nu} in the position representation and in the $h\rightarrow\delta$ limit is an ideal starting point to investigate the relationship of $\iket{\CX^{(\nu)}}{\hat{R}}$ with other density-based frameworks. 
These expressions reveal the connection between these invariant kets and several popular fingerprints designed to capture pair and 3-body interactions. 
The link between $\bra{\alpha r}\ket{\xj^{(1)}}$ and the pair distribution function~\cite{vonl+15ijqc} is obvious. Behler-Parrinello symmetry functions, and similar weighed averages of $n$-body correlations, can be seen as projections of the $SO(3)$ invariant ket over suitable test functions $G$. For instance, for a 2-body symmetry function $G_2(r)$ one has
\begin{equation}
\bra{\alpha \beta G_2}\ket{\xj} = 
\bra{\alpha}\ket{\alpha_j} \int \dint r\, G_2(r) r \bra{\beta r}\ket{\xj^{(1)}}_{\hat{R},h\rightarrow \delta},\label{eq:bpsf-g2}
\end{equation}
and an analogous expression can be written for a 3-body symmetry function  $G_3(r,r',\omega)$.
Expressions similar to \cref{eq:delta-12} can be obtained by inserting into \cref{eq:so3-realspace} Gaussians, or alternative basis functions.  
The relationship to other density-based representations, such as those discussed in Refs.~\citenum{Faber2018,Zhang2018} is less transparent, but several of the essential ingredients -- such as scaling functions that modulate geometric and chemical correlations -- can be introduced in terms of appropriate choices of the $\hat{U}$ operators, as we will discuss in the next section.

\subsection{Smooth Overlap of Atomic Positions}
\label{subsec:soap}
We have left as a last example a discussion of the connection between the symmetrized ket and the  smooth overlap of atomic positions (SOAP) power spectrum.\cite{bart+13prb,de+16pccp}
In fact, if we take as we did before  $\hat{U}_{\aleph}=1$ and  $\ket{\xj^{1}} = \ket{\xj^{2}} = \dots \ket{\xj^{\nu}}$ in Eq.~\eqref{eq:raw-nu}, the SOAP power spectrum is nothing but an alternative representation of $\iket{\xj^{(2)}}{\hat{R}}$.
To see how, one can start by expanding the translationally-invariant environmental ket \cref{eq:desc_in_pos} in a basis of orthonormal radial basis functions $R_{n}(r)$ and spherical harmonics $Y_{m}^{l}(\hat{\br})$, 
\begin{equation}
    \bra{\alpha nlm}\ket{\xj} = \int \dbr \, R_{n}(r)Y_{m}^{l}(\hat{\br})\bra{\alpha\br}\ket{\xj}.
    \label{eq:anlm-xj}
\end{equation}
Using a basis of spherical harmonics is extremely useful and practical because they block diagonalize the angular momentum operator (and thus the rotation operator), which allows for explicit integration over the rotation group in \cref{eq:desc_in_pos}. For $\nu=1$, this leads to the following feature vector,
\begin{equation}
\bra{\alpha n}\ket{{\xj}^{(1)}}_\Rhat \propto \bra{\alpha n 0 0}\ket{\xj}.
\end{equation}
For $\nu=2$, the feature vector corresponds to the SOAP power spectrum,
\begin{equation}
    \bra{\alpha n\alpha' n'l}\ket{\xj^{(2)}}_\Rhat \propto \frac{1}{\sqrt{2l+1}} \sum_{m} \bra{\alpha nlm}\ket{\xj}^{\star}
    \bra{\alpha' n'lm}\ket{\xj}.
\end{equation}
For $\nu=3$ the representation corresponds to the bispectrum~\cite{Bartok2013},
\begin{equation}
\begin{split}
&\bra{\alpha_1 n_1 l_1 \alpha_2 n_2 l_2 \alpha n l}\ket{\xj^{(3)}}_\Rhat  \propto 
\frac{1}{\sqrt{2l + 1}}
\sum_{m m_1 m_2} \bra{\xj}\ket{\alpha n l m} \\
\times&\bra{\alpha_1 n_1 l_1 m_1}\ket{\xj}
\bra{\alpha_2 n_2 l_2 m_2}\ket{\xj} \bra{l_1\, m_1\, l_2\, m_2}\ket{l\, m}.
\end{split}
\label{eq:bispectrum}
\end{equation}

where $ \bra{l_1\, m_1\, l_2\, m_2}\ket{l\, m}$ is a Clebsch-Gordan coefficient. The bispectrum is used as a four-body feature vector in SOAP and in Spectral Neighbor Analysis Potentials (SNAP), where its high resolution is exploited to construct accurate interatomic potentials through linear regression~\cite{Thompson2015}.

Seen in the light of the present formalism, the remarkable fact that the SOAP kernel (\cref{eq:haar-rotation-kernel} with densities written as a sum of Gaussians) can be expressed as an explicit scalar product between vectors, representing a truncated expansion of the power spectrum, emerges as a natural consequence of the definition of the kernel as the scalar product between invariant kets.
It should also be noted that in practical applications of SOAP the kernels are often  (but not always) normalized and raised to an integer power $\zeta$, which corresponds to taking a tensor product of the kets and introduces a many-body character in the model built on such kernels.

\begin{figure}
    \centering
    \includegraphics[width=0.9\columnwidth]{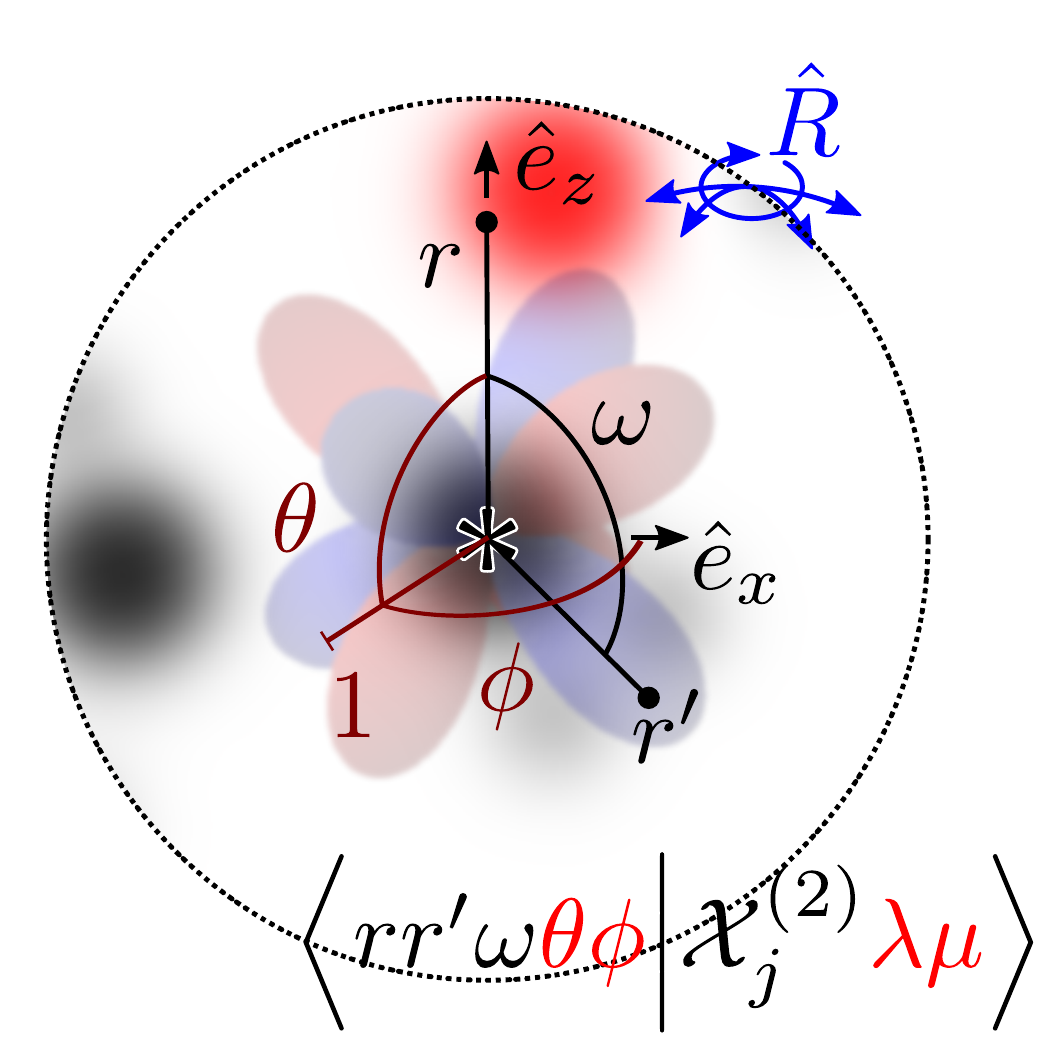}
    \caption{Schematic representation of the construction of a real-space representation of a tensorial ket associated with a $\lambda$-SOAP kernel. The (smooth) atom density is evaluated at two points corresponding to a stencil $(r,r',\omega)$, and the spherical harmonic $Y^\lambda_\mu$ is evaluated at the angles $(\theta,\phi)$, relative to the reference frame that is used to describe the stencil.}
    \label{fig:lambda-ket}
\end{figure}

\subsection{Tensorial Smooth Overlap of Atomic Positions ($\lambda$-SOAP)}

The feature vectors that appear in the tensorial extension of SOAP~\cite{gris+18prl} are of the form in \cref{eq:raw-nu}, with $\hat{U}_{\aleph} = \hat{I}$ for $\aleph = 1, 2,\dots, \nu+1$, $\ket{\mathcal{X}_{j}^{\aleph}} = \ket{\mathcal{X}_{j}}$ for $\aleph = 1, \dots, \nu$ and $\ket{\mathcal{X}_{j}^{\nu+1}} = \ket{\lambda\mu}$, where $\ket{\lambda\mu}$ is an angular momentum ket:
\begin{equation}
\iket{\CX^{(\nu)}{\lambda\mu}}{\hat{R}}  =
\int \dint\hat{R} \, \hat{R} \ket{\lambda\mu} \prod_{\aleph=1}^{\nu} \, \otimes \, \hat{R} \ket{\xj}.
\end{equation}  
The ket is rotationally invariant,
\begin{equation}
    \left[\prod_{\aleph=1}^{\nu+1} \, \otimes \, \hat{R}'\right] 
    \iket{\CX^{(\nu)}{\lambda\mu}}{\hat{R}}
    = 
    \iket{\CX^{(\nu)}{\lambda\mu}}{\hat{R}},
\end{equation}
but not in the subspace that describes the atomic environments,
\begin{equation}
 \left[ \hat{I} \otimes\prod_{\aleph=1}^{\nu} \, \otimes \, \hat{R}'\right] 
 \iket{\CX^{(\nu)}{\lambda\mu}}{\hat{R}}
 \ne 
 \iket{\CX^{(\nu)}{\lambda\mu}}{\hat{R}}.
\end{equation}
The inner product between two of these vectors is easily shown to be
\begin{equation}
    \bra{\CX^{(\nu)}_j{\lambda \mu}}\ket{\CX_k^{(\nu)}{\lambda'\mu'}}_\Rhat
    =
    \delta_{\lambda \lambda'} \int d\hat{R} \, D^{\lambda}_{\mu\mu'}(\hat{R})
    \, \big|\bra{\mathcal{X}_{j}}\hat{R}\ket{\mathcal{X}_{k}}\big|^{\nu},
\end{equation}
which agrees with the usual definition of the $\lambda$-SOAP kernel,
\begin{equation}
    \bra{\CX_j^{(\nu)}{\lambda\mu}}\ket{\CX_k^{(\nu)}{\lambda\mu'}}_\Rhat
    = k^{\lambda}_{\mu\mu'}(\mathcal{X}_j, \mathcal{X}_k).
\end{equation}
While $\ket{\CX_j^{(\nu)}{\lambda\mu}}_\Rhat$ can be represented very effectively using a spherical-harmonics expansion of the atom density~\cite{gris+18prl}, it is also possible to express it in terms of a real-space basis. Following arguments similar to those used to derive \cref{eq:bispectrum-realspace}, one can see that in this form the tensorial ket corresponds to the evaluation of a three-body correlation function of the atom density, multiplied by a spherical harmonic of appropriate order computed in the reference frame of the $(r,r',\omega)$ stencil (see \cref{fig:lambda-ket}). 

\begin{figure*}[bhtp]
\centering
\includegraphics[width=0.9\linewidth]{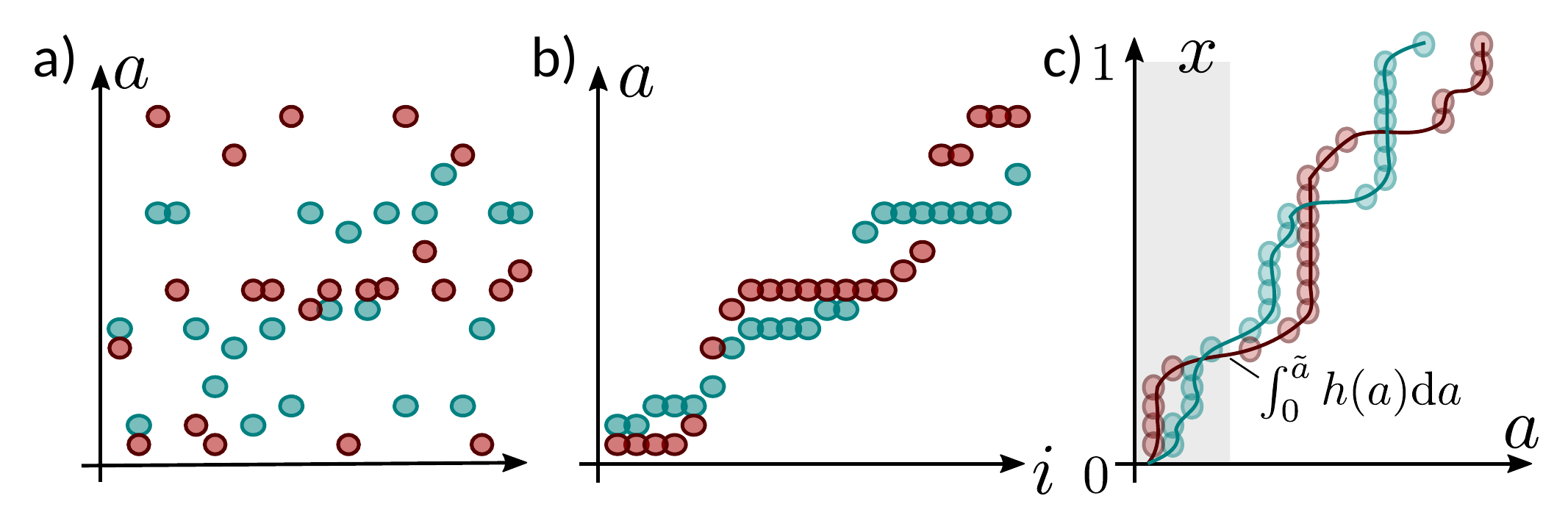}
\caption{
(a) Permutation-variant structural descriptors can be stored in a vector to be used as an atomic-scale representation. (b) Sorting this vector makes it permutationally invariant.
(c) It is easy to see how the sorted vector relates to the cumulative distribution function associated with the histogram of the values of the structural features. 
}
\end{figure*}

Taking tensor products of $\iket{\CX^{(\nu)}{\lambda\mu}}{\hat{R}}$ with itself increases the order of body correlations that are explicitly included in the feature vector, but destroys the required symmetry properties. Instead, one can take tensor products with $\lambda=0$ kets, which are rotationally invariant in the subspace that describes the atomic environments while preserving the desired symmetry of the representation, e.g.
\begin{equation}
     \iket{\CX^{(\nu)}{\lambda\mu}}{\hat{R}} \prod_{k = 1}^{\zeta-1} \otimes
     \ket{\mathcal{X}^{(\nu)}}_\Rhat \to \iket{\CX^{(\nu)}{\lambda\mu}}{\hat{R}}^{(\zeta)}.
\end{equation}
This procedure has been found effective in practice for increasing the order of body correlations in tensorial SOAP~\cite{gris+18acscs,wilk+19pnas}.

\subsection{Distributions vs sorted vectors}

It is worth making some further considerations that extend somewhat the generality of this construction to include representations that are \emph{not} based explicitly on atom densities. 
Many approaches in the literature rely on computing quantities that are not permutationally invariant \textit{per se}, for instance the elements of the matrix of pair distances between atoms~\cite{piet-mart15jcp}, transformed elementwise by some function~\cite{rupp+12prl}, or the eigenvalues of such matrices~\cite{sade+13jcp}.
In order to make these representations invariant to atom permutations, one often proceeds to sort these sets of items, and uses the Euclidean distance between the sorted vectors as the building block of kernels or other statistical learning frameworks. 

In fact, it is easy to see that given a set of elements $\left\{a_i \in \mathbb{R}\right\}$, the sorted list contains the same amount of information as the histogram of the elements $h(a)$. 
Scaling the index of the sorted items by the total number of items $N$, and considering the limit in which one can consider $x=i/N$ as a continuous index, one sees that $x(\tilde{a})$ counts the fraction of entries that are smaller than $\tilde{a}$, that is
\begin{equation}
x(\tilde{a})=\int_{-\infty}^{\tilde{a}} \textrm{d}a \, h(a).
\end{equation}
It follows that $a(x)$, which is a continuous representation of the vector of sorted distances, is just the inverse cumulative distribution function (iCDF) associated with $h(x)$. 
The Euclidean distance between two vectors of sorted elements is proportional to the $\mathcal{L}^2$ norm of the difference between the iCDF of the histograms associated with the two sets. Interestingly, if one considers the $\mathcal{L}^1$ norm, the distance between the sorted vectors corresponds to the earth mover's distance\cite{Panaretos2019} between two distributions in one dimension.
The connection between different density-based representations is more direct than that which can be established between density-based and sorted-vector descriptions -- also given that the relation between atom positions and the permutation variant items might be far from trivial, e.g. when the representation involves the eigenvalues of an overlap matrix. However, the argument we present here highlights the fact that incorporating physical symmetries in the description of atomistic systems leads to representations that contain essentially the same information. 

\section{Generalized invariant density representations}

The formalism we have introduced in the previous section provides an elegant framework to construct a rotationally-invariant representation of the atomic density that can be used for machine-learning purposes.  
While the formalism provides a complete description of structural correlations of a given order within an atomic environment, the quality and the computational cost of the regression scheme can be improved substantially in practice by transforming the representation so that it incorporates some degree of chemical intuition. 
For instance, the combination of multiple kernels corresponding to different interatomic distances has been shown to improve the quality of the ML model\cite{bart+17sa}. Likewise, a scaling of the weights of different atomic distances within an environment has been shown to be beneficial when using ML to predict atomic-scale properties~\cite{Huang2016,Faber2018}.

We will discuss how many of these modifications can be incorporated in the  through inclusion of a rotationally-invariant Hermitian operator $\hat{U} = \hat{U}_{1} = \hat{U}_{2} = \dots$ (as introduced earlier) that leads to coupling of the geometric and elemental components of the translationally-invariant atom-centered ket $\ket{\xj}$.
For concreteness, and to provide a formulation that can be directly applied to an existing framework, we discuss $\hat{U}$ written in the orthonormal basis of radial functions and spherical harmonics $\{\ket{\alpha nlm}\}$, that correspond to the SOAP power spectrum. 

The requirement that  $\hat{U}$ is rotationally-invariant (and thus commutes with an arbitrary rotation operator) means that it must have  the following form 
\begin{equation}
    \bra{\alpha nlm}\hat{U}\ket{\alpha' n'l'm'} = \, \delta_{ll'} \delta_{mm'}  \bra{\alpha
    n l}\hat{U}\ket{\alpha' n' l'}.
\label{eq:wop-nalpha}    
\end{equation}
Eq.~\eqref{eq:wop-nalpha} is the most general form compatible with  $SO(3)$ symmetry, and can be seen as a way to introduce correlations between different radial and elemental components of the features, and to weight the contribution from different angular channels. 

\subsection{Low-rank expansion of the $\hat{U}$ operator}
\label{subsec:compression}

Since $\hat{U}$ is Hermitian, it can be diagonalized and expressed in the orthogonal basis of its
eigenkets $\{\ket{\overline{J}}\}$,
\begin{equation}
\hat{U} = \sum_{J} \ket{\overline{J}} U_{J}\bra{\overline{J}}.
\end{equation}
Taking $U_{J}\bra{\overline{J}} \to \bra{J}$ allows us to express $\hat{U}$ as $\hat{U} = \sum_{J} \ket{\overline{J}} \bra{J}$.

The transformed SO(3) vector components can be written in terms of the components of $\ket{J}$ in the chemical basis, $u_{J\alpha n l} = \bra{J}\ket{\alpha n l}$. This yields
\begin{equation}
    \begin{split}
    \bra{JJ'}\ket{\xj^{(2)}}_\Rhat =&  \sum_{\alpha \alpha' n n' l} u_{J\alpha n l}u_{J'\alpha'n'l} \\
  & \times \sum_{m}  \bra{\alpha nlm}\ket{\xj}^{\star} \bra{\alpha' n'lm}\ket{\xj}.
    \end{split}
    \label{eq:lowrank-feat}
\end{equation}
By choosing a low-rank expansion of $\hat{U}$ one can greatly reduce the dimensionality of the $SO(3)$ fingerprint vector, similarly to what was done in Ref.~\citenum{imba+18jcp} applying standard sparse decomposition techniques to the $SO(3)$ fingerprints.

A possible approach is to determine this low-rank approximation based on the correlations found between environments that are part of the data set. 
For a given $l$, consider the spherically-symmetric covariance matrix between the features of the expanded atomic density,\footnote{Note that, apart from a $l$-dependent scaling, the covariance matrix is just the average of the SOAP power spectrum over the training set. }
\begin{equation}
\begin{split}
    C_{\alpha n\alpha'n'}^{(l)} = & \frac{1}{N}\sum_{j} \sum_{m} \bra{\alpha nlm}\ket{\mathcal{X}_{j}}^\star \bra{\mathcal{X}_{j}}\ket{\alpha'n'lm} \\
    = & \frac{\sqrt{2l+1}}{N}\sum_{j} \bra{\alpha n \alpha' n' l}\ket{\CX_j^{(2)}}_\Rhat.
\end{split}
    \label{eq:cov}
\end{equation}
The eigenvectors of $\mathbf{C}^{(l)}$,  $\bm{v}_{J}^{(l)}$,  can then be used as $u_{J \alpha n l}$ in Eq.~\eqref{eq:lowrank-feat}. It is easy to see that this transformation identifies components of the data that are linearly independent within the training set, and have a spread that is equal to the corresponding eigenvalues $\lambda_J^{(l)}$. The feature space can then be compressed by only retaining a certain number of components $n_J$ that could be determined using the magnitude of the associated eigenvalues.

\subsection{Radially-scaled kernels}
\label{sub:radial}
In a system with relatively uniform atom density, the overlap between environments $\bra{\xj}\ket{\xk}$ is dominated by the region farthest from the center.  This could be regarded as rather unphysical, since interactions between atoms decay with distance and the closest atoms should therefore give the most significant contribution to properties, which is reflected in the observation that multi-scale kernels tend to perform best when very low weights are assigned to the long-range kernels~\cite{Bartok2017,paru+18ncomm,wilk+19pnas}.  This effect can be counteracted by multiplying the atomic probability amplitude Eq.~\eqref{eq:r-xj} with a radial scaling $u(r)$,
\begin{equation}
    \bra{\alpha\br}\hat{U}\ket{\xj} = u(r) \psi_{\xj}^\alpha(\br).
    \label{eq:r-U-xj}
\end{equation}

In the context of the  SOAP power spectrum, this change can be represented in terms of a $\hat{U}$ operator that reads
\begin{equation}
\bra{n}\hat{U}\ket{n'} =  \int \dint r\, r^2  R_n(r)R_{n'}(r) u(r),
\label{eq:wop-radial}
\end{equation}
since an operator that scales states in the position representation must be diagonal in it,
\begin{equation}
    \bra{r}\hat{U}\ket{r'} = \delta(r - r') u(r),
\end{equation}
and its matrix elements in the basis of radial basis functions are
\begin{equation}
    \bra{n}\hat{U}\ket{n'} = \int dr \int dr' r^{2} R_{n}(r) R_{n'}(r') \delta(r - r') u(r),
\end{equation}
which reduces to \cref{eq:wop-radial}.

Radial scaling in the form of Eq.~\eqref{eq:r-U-xj} can be approximated, when using narrow atom-centered functions, with 
$\sum_i u(r_{ij}) f_c(r_{ij}) h(\br -\br_{ij})$, where we also consider for simplicity the case with a single species~\cite{will+18pccp}. 
Besides the fact that it is simpler to implement this form of scaling in an existing code, this approximation also makes apparent the connection between the general density-based framework we introduce here and the descriptors of Ref.~\citenum{Faber2018}.
When $h$ is taken to be a Gaussian function, the weight on the central atom is set to zero and one considers the two-body invariant representations, this ansatz is essentially equivalent to the two-body features in Ref.~\citenum{Faber2018}: 
\begin{equation}
\begin{split}
\mel{r}{\hat{U}}{\xj^{(1)}}_\Rhat =&
\sum_{i\ne j} u(r_{ij}) \frac{\sqrt{2\pi}}{r_{ij} \sigma} \left[
e^{-\frac{(r-r_{ij})^2}{2\sigma^2}} - e^{-\frac{(r+r_{ij})^2}{2\sigma^2}}\right]\\
\approx & \sum_{i\ne j} u(r_{ij}) \frac{\sqrt{2\pi}}{r_{ij}} e^{-\frac{(r-r_{ij})^2}{2\sigma^2}}.
\end{split}
\end{equation}

\subsection{Alchemical kernels}

In the presence of multiple species, one could make the scaling element dependent, or devise a more complex operator that couples different channels of different species.  As a first test of the generalization of SOAP in the presence of multiple elements,  we consider an operator in the form

\begin{equation}
    \bra{\alpha nlm}\hat{U}\ket{\alpha' n'l'm'} = \, \delta_{ll'} \delta_{mm'} \delta_{nn'} 
    \bra{\alpha}\hat{U}\ket{\alpha'},
\end{equation}
which ignores couplings between the structure of an environment and the elements within it.  
One can always write a low-rank expansion of the operator, $\hat{U}\approx\sum_{J\alpha} \ket{\overline{J}} u_{J\alpha} \bra{\alpha}$, which allows one to write 
\begin{equation}
\hat{U}\otimes\hat{U} \iket{\xj^{(2)}}{\hat{R}}=\sum_{\alpha\alpha'}
\ket{\overline{JJ'}}u_{J\alpha} u_{J'\alpha'} \ibraket{\alpha\alpha'}{\xj^{(2)}}{\hat{R}}.
\end{equation}
In the context of SOAP, one can define the projections of the power spectrum in this ``alchemical basis'',
\begin{equation}
    \begin{split}
    \bra{JnJ'n'l}\ket{\xj^{(2)}}_\Rhat =& \sum_{\alpha \alpha'} u_{J\alpha}u_{J'\alpha'}
   \sum_{m} \bra{\alpha nlm}\ket{\xj} \\
    & \times \bra{\alpha' n'lm}\ket{\xj},
    \label{eq:alchemy-feat}
    \end{split}
\end{equation}
which was shown in Ref.~\citenum{will+18pccp} to yield a substantial improvement in the learning efficiency in the presence of many chemical elements, and to result in a low-dimensional representation of elemental space that shares some similarities with the grouping found in the periodic table of the elements. 

One can see the relationship between these ``alchemical features'' and previous attempts to incorporate cross-species correlations through the generalized SOAP environmental kernel,
\begin{equation}
    \begin{split}
    \label{eq:alchemy-kernel}
     \int \drhat \left| \bra{\xj}\hat{U}^{\dagger}\hat{U}\hat{R}\ket{\xk} \right|^{2} = \\
    \sum_{JnJ'n'l} {}_\Rhat{\bra{\xj^{(2)}}\ket{JnJ'n'l}} \bra{JnJ'n'l}\ket{\xk^{(2)}}_\Rhat .
    \end{split}
\end{equation}
By writing out explicitly this inner product in terms of the full power spectrum elements $\ibraket{\alpha n \alpha' n' l}{\xj^{(2)}}{\hat{R}}$ one can see that the matrix elements $\bra{\alpha} \hat{U}^{\dagger}\hat{U}\ket{\alpha'}$ are nothing but the elements of the alchemical kernel $\kappa_{\alpha\alpha'}$ that was introduced in Ref.~\citenum{De2016}, where it was shown that taking $\kappa_{\alpha\alpha'} \ne \delta_{\alpha\alpha'}$ can improve property predictions with kernel ridge regression.\cite{De2016,Bartok2017}
Off-diagonal couplings between chemical elements have also been used in other representations, including those of Ref.~\citenum{Faber2018}.

The expression in terms of reduced features \cref{eq:alchemy-feat} is, however, more efficient to compute and clarifies how this approach enables the introduction of correlations between elements, as well as reduction of the space dimensionality. The full SOAP feature vector contains a number of components that is proportional to the square of the number of present species $n_\text{sp}$, while limiting to a number $d_J\ll n_\text{sp}$ of basis kets reduces the dimensionality of the feature vector by a factor $(n_\text{sp}/d_J)^2$. 

Note that one does not even need to compute all the elements in the $\ket{\alpha nlm}$ expansion of the density, since the alchemical projection can be brought down to the level of the atom density, which can be defined for $d_J$ chemical ``channels'' rather than for each element separately,
\begin{equation}
\bra{J\br}\ket{\xj} = \sum_\alpha u_{J\alpha} \psi_{\xj}^\alpha(\br).
\label{eq:alchemy-density}
\end{equation}
Density-based representations that assign a weight to each species have been explored as means to reduce the complexity of ML representations in cases where many elements are present simultaneously~\cite{nong+17prb,gast+18jcp,Huo2017}, which correspond essentially to the case with $d_J=1$.
For instance, the \emph{compositional descriptor} of Ref.~\citenum{nong+17prb} is equivalent to Eq.~\eqref{eq:bpsf-g2} computed on a single invariant density,
\begin{equation}
\ibraket{r}{\xj^{(1)}}{\hat{R}}=\sum_i u_{\alpha_i} \delta(r-r_{ij}) f_c(r_{ij}),
\end{equation}
where the weights of different species are rather arbitrarily set to be $u_\alpha=0,\pm 1, \pm 2\ldots$.
The more general formulation in Eqs.~\eqref{eq:alchemy-feat}-\eqref{eq:alchemy-density} provides a way to alter the dimensionality of the representation, and to optimize the projections to obtain the most efficient features for a given regression problem.

\subsection{Non-factorizable operators}

In order to relate \cref{eq:raw-nu} to other density-based representations that involve more complicated scaling functions of the internal coordinates, it is necessary to introduce a further linear transformation $\hat{U}^{(\nu)}$ which does not factorize into components that act independently on each term in the $\nu$-order tensor product. Such an operator must be chosen with care to ensure that it is rotationally-invariant, otherwise the rotational-invariance of the transformed ket will be lost. As far as the $\nu=2$ rotationally-invariant kets are concerned, a generic operator is completely determined by its action on the basis vectors $\{\ket{\alpha \mathbf{r} \alpha' \mathbf{r}'}\}$. Rotationally-invariant operators must act on $\ket{\alpha \hat{R}\mathbf{r} \alpha' \hat{R}\mathbf{r}'}$ in the same was as on $\ket{\alpha \mathbf{r} \alpha' \mathbf{r}'}$, followed by the rotation $\hat{R}$. The upshot of this observation is
\begin{equation}
    \bra{\alpha \mathbf{r}_{1} \alpha \mathbf{r}_{1}'}\hat{U}^{(2)} \ket{\beta \mathbf{r}_{2} \alpha \mathbf{r}_{2}'}
    = \bra{\alpha r_{1} \alpha' r_{1}' \omega_{1}} \hat{U}^{(2)} \ket{\beta r_{2} \beta' r_{2}' \omega_{2}},
\end{equation}
i.e. any non-internal coordinate must be cyclic. If a distance and angle-based scaling is required, then the operator is diagonal,
\begin{equation}
\begin{split}
    \bra{\alpha \mathbf{r}_{1} \alpha \mathbf{r}_{1}'}\hat{U}^{(2)} \ket{\beta \mathbf{r}_{2} \alpha \mathbf{r}_{2}'}
    =& \, \delta_{\alpha\beta}\delta_{\alpha'\beta'} \delta(r_{1} - r_{2}) \delta(r_{1}' - r_{2}')  \\
    & \times \delta(\omega_{1} - \omega_{2}) u(\alpha, r_{1}, \alpha', r_{1}', \omega_{1}).
\end{split}
\end{equation}
For example, the scaling function in the three-body descriptor in Ref.~\citenum{Faber2018} corresponds to the following choice for $u(r_{1}, r_{2}, \omega)$,
\begin{equation}
     u(r_{1}, r_{2}, \omega_{1}) = \frac{1 - 3 \omega_{1}\omega_{2}\omega_{3}}{(r_{1}r_{2}r_{3})^{n}},
\end{equation}
where $r_{3}^{2} = r_{1}^{2} + r_{2}^{2} - 2r_{1}r_{2}\omega_{1}$, $\omega_{2} = (r_{1}^{2} - r_{2}^{2} - r_{3}^{2})/ 2r_{2}r_{3}$, $\omega_{3} = (r_{2}^{2} - r_{1}^{2} - r_{3}^{2})/ 2r_{1}r_{3}$ and $n$ is an adjustable parameter. Faber \textit{et al}. do not specify a scaling function for four-body and higher-body descriptors, but the analysis presented here clearly extends to any hypothetical scaling function that involves the internal coordinates of a collection of $\nu+1$ positions.

Starting from the SOAP power spectrum, one can exploit the fact that each component is separately symmetry invariant. It is then possible to introduce an arbitrary linear operator coupling the $\ket{\alpha n \alpha' n'l}$ components, $\bra{\alpha n_{1} \alpha' n_{1}'l_{1}}\hat{U} \ket{\beta n_{2}\beta' n_{2}'l_{2}}$. 
Being a linear operation, this transformation amounts to a change of regularization for the ridge regression problem, and is most useful if applied to reduce the dimensionality of the feature vectors. 
This can be done e.g. by finding the principal components of the covariance matrix of the SOAP power spectrum or -- as done in Ref.~\citenum{imba+18jcp} -- by a sparse decomposition that singles out a subset of the components that suffice to obtain a thorough description of the relevant structures.
This corresponds to the contracted representation
\begin{align}
\ibraket{J}{\xj^{(2)}}{\hat{R}} = \sum_{Jk} u_{Jk} \ibraket{\alpha_k n_k\alpha_k' n_k' l_k}{\xj^{(2)}}{\hat{R}},
\label{eq:giulio-svd}
\end{align}
where $k$ runs over the set of selected components,\footnote{The sparsification could be also represented explicitly by a $\hat{U} $ operator, that zeroes out all of the unnecessary components.} which can be determined with different schemes, from CUR~\cite{Mahoney2009} to farthest point sampling~\cite{rose+77siam,ceri+13jctc}. The coefficients $u_{Jk}$ are the elements of a square matrix that ensures the contracted vectors in Eq.~\eqref{eq:giulio-svd} generate a kernel that is as close as possible to the full kernel.

\subsection{Optimization of the density representation}

\label{subsec:opt_rep}

The optimization of the $\hat{U}$ operator in its more general form (see \cref{eq:wop-nalpha}) involves a large number of parameters, leading to a very concrete risk of overfitting. This is exacerbated by the fact that the feature vector is then used as the input for regression, and one has to balance the amount of data used to optimize the elements of $\hat{U}$ and that used for the training of the ridge regression model.
The simplest approach to reduce the optimization of $\hat{U}$ to a small number of free parameters uses the compression method discussed in \cref{subsec:compression} to identify the most important combinations of $\bra{\alpha n l m}\ket{\CX_j}$ components that are linearly independent for the data at hand. 
We would like to be able to optimize $\bu$ based on the correlations found between environments that are part of the training set. The idea is that further optimization using target properties will be less likely to overfit after this dimensionality reduction.

Another possible use of the principal-component representation of  $\bra{\alpha n l m}\ket{\CX_j}$ is to obtain a simpler ansatz to further optimize the $\hat{U}$ operator. For instance, one could combine linearly the different components using the $\hat{U}$ operator defined in \cref{eq:cov}
\begin{equation}
    \bra{IJlm}\ket{\mathcal{X}_{j}} = \sum_{I'\alpha J' n} f^{(l)}_{II'JJ'} u^{(l)}_{I'\alpha J' n}\bra{\alpha nlm}\ket{\mathcal{X}_{j}},
    \label{eq:sh_contraction_scaled}
\end{equation}
where the scaling coefficients $f^{(l)}_{II'JJ'}$ are determined so as to make the representation better suited to build a regression model for the target property $y$. 
A systematic exploration of the different possibilities, as well as their benchmarking on different regression problems, is left for future work.

\section{Conclusions}

We have introduced a general formulation of the problem of representing atomic structures in terms of a (smooth) atom density, which is independent on the basis that is used to expand it.
Starting from a representation of a 3D structure in terms of a superposition of atom-centered functions decorated with elemental kets, we introduce symmetries by formally averaging the feature vectors over the continuous translation and rotation groups. 
The averaging removes information, but a complete, unique description can be retained by taking tensor products of the ket before computing the integral. Different representations, capturing varying amounts of inter-atomic correlations, can be obtained depending on the combination of tensor products and symmetrized averages.

The framework we introduced provides a unified picture of density-based representations for machine learning of atomic-scale properties, with several popular frameworks emerging by taking different limits, or using specific basis sets to represent the abstract invariant kets. 
In particular, using a basis of radial functions and spherical harmonics shows clearly the 1:1 mapping between the symmetrized kets and different flavors of the SOAP kernel.
Even alternative schemes that start from rotation and translation-invariant internal coordinates and proceed to ensure permutation invariance appear to contain comparable information.
Physically-motivated extensions of existing frameworks, relying also on real-space formulations that are more directly related to body-order expansions and to atom correlation functions that appear in classical density functional theory, might provide some leeway to improve the performance of a representation, as shown for instance in Ref.~\citenum{will+18pccp}. 
We discussed how several modifications and optimizations can be introduced in terms of operators that couple and scale different channels of the representation, focusing in particular on the  SOAP power spectrum representation. 
Such modifications can be advantageous as they allow for a reduction in the dimensionality of the problem, and make it possible to incorporate different kinds of chemical insights -- effectively generating  several ``views'' of a material that emphasize different kinds of structural and chemical correlations. 
Seeing many different approaches as alternative implementations of the same family of representations might help coordinate efforts in optimizing the computational efficiency and regression accuracy of  density-based models of atomic structures. 

\section*{Acknowledgements}

The Authors would like to thank G\'abor Cs\'anyi, Bingqing Cheng and Andrea Grisafi  for discussion and comments during the preparation of this work. MC and MJW were supported by the European Research Council under the European Union's Horizon 2020 research and innovation programme (grant agreement no. 677013-HBMAP). FM was supported by the the NCCR MARVEL, funded by the Swiss National Science Foundation.

\bibliographystyle{bibliography/aip}

\begin{thebibliography}{10}

\bibitem{Ramakrishnan2015}
R.~Ramakrishnan and O.~A. von Lilienfeld,
\newblock  {\bf 69}, 182 (2015).

\bibitem{Faber2016}
F.~A. Faber, A.~Lindmaa, O.~A. von Lilienfeld, and R.~Armiento,
\newblock Phys. Rev. Lett. {\bf 117}, 135502 (2016).

\bibitem{Bartok2017}
A.~P. Bart{\'{o}}k et~al.,
\newblock Sci. Adv. {\bf 3}, e1701816 (2017).

\bibitem{Montavon2013}
G.~Montavon et~al.,
\newblock New J. Phys. {\bf 15}, 095003 (2013).

\bibitem{Mitchell2014}
J.~B.~O. Mitchell,
\newblock Wiley Interdisciplinary Reviews: Computational Molecular Science {\bf
  4}, 468 (2014).

\bibitem{Carrete2014}
J.~Carrete, W.~Li, N.~Mingo, S.~Wang, and S.~Curtarolo,
\newblock Physical Review X {\bf 4}, 011019 (2014).

\bibitem{Ward2016}
L.~Ward, A.~Agrawal, A.~Choudhary, and C.~Wolverton,
\newblock npj Computational Materials {\bf 2}, 16028 (2016).

\bibitem{behl-parr07prl}
J.~Behler and M.~Parrinello,
\newblock Phys. Rev. Lett. {\bf 98}, 146401 (2007).

\bibitem{Szlachta2014}
W.~J. Szlachta, A.~P. Bart{\'{o}}k, and G.~Cs{\'{a}}nyi,
\newblock Phys. Rev. B {\bf 90}, 104108 (2014).

\bibitem{Bartok2010}
A.~P. Bart{\'{o}}k, M.~C. Payne, R.~Kondor, and G.~Cs{\'{a}}nyi,
\newblock Phys. Rev. Lett. {\bf 104}, 136403 (2010).

\bibitem{Morawietz2012}
T.~Morawietz, V.~Sharma, and J.~Behler,
\newblock Journal of Chemical Physics {\bf 136}, 064103 (2012).

\bibitem{Kobayashi2017}
R.~Kobayashi, D.~Giofr{\'{e}}, T.~Junge, M.~Ceriotti, and W.~A. Curtin,
\newblock Phys. Rev. Mater. {\bf 1}, 053604 (2017).

\bibitem{Smith2017}
J.~S. Smith, O.~Isayev, and A.~E. Roitberg,
\newblock Chem. Sci. {\bf 8}, 3192 (2017).

\bibitem{Khorshidi2016}
A.~Khorshidi and A.~A. Peterson,
\newblock Computer Physics Communications {\bf 207}, 310 (2016).

\bibitem{Gastegger2017}
M.~Gastegger, J.~Behler, and P.~Marquetand,
\newblock Chem. Sci. {\bf 8}, 6924 (2017).

\bibitem{Glielmo2018}
A.~Glielmo, C.~Zeni, and A.~{De Vita},
\newblock Physical Review B {\bf 97} (2018).

\bibitem{Chmiela2017}
S.~Chmiela et~al.,
\newblock Science Advances {\bf 3}, e1603015 (2017).

\bibitem{Yao2017}
K.~Yao, J.~E. Herr, and J.~Parkhill,
\newblock Journal of Chemical Physics {\bf 146}, 014106 (2017).

\bibitem{Bartok2013}
A.~P. Bart{\'{o}}k, R.~Kondor, and G.~Cs{\'{a}}nyi,
\newblock Phys. Rev. B {\bf 87}, 184115 (2013).

\bibitem{Rupp2012}
M.~Rupp, A.~Tkatchenko, K.-R. M{\"{u}}ller, and O.~A. von Lilienfeld,
\newblock Physical Review Letters {\bf 108}, 058301 (2011).

\bibitem{Barker2017}
J.~Barker, J.~Bulin, J.~Hamaekers, and S.~Mathias,
\newblock {LC-GAP: Localized coulomb descriptors for the gaussian approximation
  potential},
\newblock in {\em Scientific Computing and Algorithms in Industrial
  Simulations: Projects and Products of Fraunhofer SCAI}, pages 25--42,
  Springer, Cham, 2017.

\bibitem{Sadeghi2013}
A.~Sadeghi et~al.,
\newblock Journal of Chemical Physics {\bf 139}, 184118 (2013).

\bibitem{Zhang2018}
L.~Zhang, J.~Han, H.~Wang, R.~Car, and E.~Weinan,
\newblock Physical Review Letters {\bf 120}, 143001 (2018).

\bibitem{braa-bowm09irpc}
B.~J. Braams and J.~M. Bowman,
\newblock International Reviews in Physical Chemistry {\bf 28}, 577 (2009).

\bibitem{Xie2010}
Z.~Xie and J.~M. Bowman,
\newblock Journal of Chemical Theory and Computation {\bf 6}, 26 (2010).

\bibitem{Jiang2013}
B.~Jiang and H.~Guo,
\newblock The Journal of Chemical Physics {\bf 054112}, 0 (2013).

\bibitem{Schutt2014}
K.~T. Sch{\"{u}}tt et~al.,
\newblock Phys. Rev. B {\bf 89}, 205118 (2014).

\bibitem{Hirn2017}
M.~Hirn, S.~Mallat, and N.~Poilvert,
\newblock Multiscale Modeling {\&} Simulation {\bf 15}, 827 (2017).

\bibitem{Eickenberg2017}
M.~Eickenberg, G.~Exarchakis, M.~Hirn, and S.~Mallat,
\newblock Advances in Neural Information Processing Systems 30 , 6522 (2017).

\bibitem{Shapeev2015}
A.~V. Shapeev,
\newblock Multiscale Modeling {\&} Simulation {\bf 14}, 1153 (2015).

\bibitem{Bartok2013a}
A.~P. Bart{\'{o}}k, M.~J. Gillan, F.~R. Manby, and G.~Cs{\'{a}}nyi,
\newblock Phys. Rev. B {\bf 88}, 054104 (2013).

\bibitem{Deringer2017}
V.~L. Deringer and G.~Cs{\'{a}}nyi,
\newblock Phys. Rev. B {\bf 95}, 094203 (2017).

\bibitem{Jinnouchi2017}
R.~Jinnouchi and R.~Asahi,
\newblock Journal of Physical Chemistry Letters {\bf 8}, 4279 (2017).

\bibitem{PhysRevB.90.104108}
W.~J. Szlachta, A.~P. Bart{\'{o}}k, and G.~Cs{\'{a}}nyi,
\newblock Physical Review B - Condensed Matter and Materials Physics {\bf 90},
  104108 (2014).

\bibitem{Kajita2017}
S.~Kajita, N.~Ohba, R.~Jinnouchi, and R.~Asahi,
\newblock Scientific Reports {\bf 7} (2017).

\bibitem{Grisafi2018}
A.~Grisafi, D.~M. Wilkins, G.~Cs{\'{a}}nyi, and M.~Ceriotti,
\newblock Phys. Rev. Lett. {\bf 120}, 036002 (2018).

\bibitem{imba+18jcp}
G.~Imbalzano et~al.,
\newblock J. Chem. Phys. {\bf 148}, 241730 (2018).

\bibitem{Rasmussen2006}
C.~E. Rasmussen and C.~K.~I. Williams,
\newblock {\em {Gaussian processes for machine learning}},
\newblock MIT Press, 2006.

\bibitem{glie+17prb}
A.~Glielmo, P.~Sollich, and A.~{De Vita},
\newblock Phys. Rev. B {\bf 95}, 214302 (2017).

\bibitem{nachbin1976haar}
L.~Nachbin,
\newblock {\em {The Haar integral}},
\newblock R. E. Krieger Pub. Co., 1976.

\bibitem{Note1}
For a generic basis function $g*g$ might be a complicated function, but when
  $g$ is a Gaussian, $g*g$ is simply a Gaussian with double the variance.

\bibitem{De2016}
S.~De, A.~P. Bart{\'{o}}k, G.~Cs{\'{a}}nyi, and M.~Ceriotti,
\newblock Phys. Chem. Chem. Phys. {\bf 18}, 13754 (2016).

\bibitem{prod-kohn05pnas}
E.~Prodan and W.~Kohn,
\newblock Proc. Natl. Acad. Sci. USA {\bf 102}, 11635 (2005).

\bibitem{yang91prl}
W.~Yang,
\newblock Phys. Rev. Lett. {\bf 66}, 1438 (1991).

\bibitem{gall-parr92prl}
G.~Galli and M.~Parrinello,
\newblock Phys. Rev. Lett. {\bf 69}, 3547 (1992).

\bibitem{goed99rmp}
S.~Goedecker,
\newblock Rev. Mod. Phys. {\bf 71}, 1085 (1999).

\bibitem{ceri+08jcp}
M.~Ceriotti, T.~D. K{\"{u}}hne, and M.~Parrinello,
\newblock J. Chem. Phys. {\bf 129}, 24707 (2008).

\bibitem{Note2}
This is a consequence of the fact that $SO(3)$ is the product of $SO(2)$ and
  $S^2$.

\bibitem{zile+18nc}
A.~Ziletti, D.~Kumar, M.~Scheffler, and L.~M. Ghiringhelli,
\newblock Nature Comm. {\bf 9}, 2775 (2018).

\bibitem{vonl+15ijqc}
O.~A. von Lilienfeld, R.~Ramakrishnan, M.~Rupp, and A.~Knoll,
\newblock International Journal of Quantum Chemistry {\bf 115}, 1084 (2015).

\bibitem{Faber2018}
F.~A. Faber, A.~S. Christensen, B.~Huang, and O.~A. {Von Lilienfeld},
\newblock Journal of Chemical Physics {\bf 148}, 241717 (2018).

\bibitem{bart+13prb}
A.~P. Bart{\'{o}}k, R.~Kondor, and G.~Cs{\'{a}}nyi,
\newblock Phys. Rev. B {\bf 87}, 184115 (2013).

\bibitem{de+16pccp}
S.~De, A.~A.~P. Bart{\'{o}}k, G.~Cs{\'{a}}nyi, and M.~Ceriotti,
\newblock Phys. Chem. Chem. Phys. {\bf 18}, 13754 (2016).

\bibitem{Thompson2015}
A.~Thompson, L.~Swiler, C.~Trott, S.~Foiles, and G.~Tucker,
\newblock J. Comput. Phys. {\bf 285}, 316 (2015).

\bibitem{gris+18prl}
A.~Grisafi, D.~D.~M. Wilkins, G.~Cs{\'{a}}nyi, and M.~Ceriotti,
\newblock Phys. Rev. Lett. {\bf 120}, 036002 (2018).

\bibitem{gris+18acscs}
A.~Grisafi et~al.,
\newblock ACS Central Science {\bf 5}, 57 (2019).

\bibitem{wilk+19pnas}
D.~M. Wilkins et~al.,
\newblock arxiv:1809.05337  (2018).

\bibitem{piet-mart15jcp}
F.~Pietrucci and R.~Martoň{\'{a}}k,
\newblock J. Chem. Phys. {\bf 142}, 104704 (2015).

\bibitem{rupp+12prl}
M.~Rupp, A.~Tkatchenko, K.-R. M{\"{u}}ller, and O.~A. von Lilienfeld,
\newblock Phys. Rev. Lett. {\bf 108}, 058301 (2012).

\bibitem{sade+13jcp}
A.~Sadeghi et~al.,
\newblock J. Chem. Phys. {\bf 139}, 184118 (2013).

\bibitem{Panaretos2019}
V.~M. Panaretos and Y.~Zemel,
\newblock Annual Review of Statistics and Its Application {\bf 6}, annurev
  (2019).

\bibitem{bart+17sa}
A.~A.~P. Bart{\'{o}}k et~al.,
\newblock Science Advances {\bf 3}, e1701816 (2017).

\bibitem{Huang2016}
B.~Huang and O.~A. von Lilienfeld,
\newblock J. Chem. Phys. {\bf 145}, 161102 (2016).

\bibitem{Note3}
Note that, apart from a $l$-dependent scaling, the covariance matrix is just
  the average of the SOAP power spectrum over the training set.

\bibitem{paru+18ncomm}
F.~M. Paruzzo et~al.,
\newblock Nature Comm. {\bf 9}, 4501 (2018).

\bibitem{will+18pccp}
M.~J. Willatt, F.~Musil, and M.~Ceriotti,
\newblock Physical Chemistry Chemical Physics {\bf 20}, 29661 (2018).

\bibitem{nong+17prb}
N.~Artrith, A.~Urban, and G.~Ceder,
\newblock Phys. Rev. B {\bf 96}, 014112 (2017).

\bibitem{gast+18jcp}
M.~Gastegger, L.~Schwiedrzik, M.~Bittermann, F.~Berzsenyi, and P.~Marquetand,
\newblock J. Chem. Phys. {\bf 148}, 241709 (2018).

\bibitem{Huo2017}
H.~Huo and M.~Rupp,
\newblock arxiv 1704.06439  (2017).

\bibitem{Note4}
The sparsification could be also represented explicitly by a $\protect
  \mathaccentV {hat}05E{U} $ operator, that zeroes out all of the unnecessary
  components.

\bibitem{Mahoney2009}
M.~W. Mahoney and P.~Drineas,
\newblock Proceedings of the National Academy of Sciences {\bf 106}, 697
  (2009).

\bibitem{rose+77siam}
D.~J. Rosenkrantz, R.~E. Stearns, and P.~M. {Lewis, II},
\newblock SIAM Journal on Computing {\bf 6}, 563 (1977).

\bibitem{ceri+13jctc}
M.~Ceriotti, G.~A. Tribello, and M.~Parrinello,
\newblock J. Chem. Theory Comput. {\bf 9}, 1521 (2013).

\end{thebibliography}

\onecolumngrid
\clearpage
\newpage

\end{document}